\documentclass[aps,prc,preprint,tightenlines,showpacs,%
amssymb,byrevtex,nofootinbib]{revtex4}  

\usepackage{epsfig,bm,dcolumn}
\begin{document}

\preprint{hep-ph/0602112}


\title{$\bm{K^*}$ photoproduction off the nucleon: $\bm{\gamma N \to K^*
\Lambda}$}


\author{Yongseok Oh}%
\email{yoh@physast.uga.edu}

\affiliation{Department of Physics and Astronomy,
University of Georgia, Athens, Georgia 30602, U.S.A.}

\author{Hungchong Kim}%
\email{hungchon@postech.ac.kr}

\affiliation{Department of Physics, Pohang University of Science and
Technology, Pohang 790-784, Korea}



\begin{abstract}

We study the photoproduction of $K^*(892)$ vector meson 
from both the charged and neutral reactions,
$\gamma p \to K^{*+} \Lambda$ and $\gamma n \to K^{*0} \Lambda$.
The production mechanisms that we consider include 
$t$-channel $K^*$, $K$,
$\kappa$ exchanges, $s$-channel nucleon diagram, and $u$-channel $\Lambda$,
$\Sigma$, $\Sigma^*$ diagrams.
These could constitute important backgrounds for future investigation
of ``missing'' resonances that can be searched for especially in these
reactions.
The $t$-channel $K$ meson exchange is found to dominate both reactions.
The total and differential cross sections are presented together with
some spin asymmetries.

\end{abstract}

\pacs{13.60.Le, 13.60.-r, 13.60.Rj}

\maketitle

\section{Introduction}

The baryon spectra predicted by some quark models anticipate much more
baryon resonances than the observed so far \cite{CR00}.
These ``missing'' resonances are expected to have rather small couplings to
the $\pi N$ channel, and 
various reaction mechanisms have been suggested to search
for those resonances.
One of them is to use photoproduction processes containing mesons other
than pion(s) in the final state.
For example, the photoproductions of $K\Lambda$ and $K\Sigma$ in the
scattering off the nucleon may give us a clue on the
existence of nucleon resonances that strongly couple to the kaon
channel~\cite{CR98b}.
Vector meson photoproduction, $\gamma N \to VN$, where $V$ stands for a
vector meson ($\rho$,$\omega$,$\phi$), may also be useful to identify
the missing resonances~\cite{OTL01}.

Recently, the interest in $K^*(892)$ vector meson photoproduction has
been grown.
This was initially triggered by the quark model which predicts 
that some nucleon resonances with higher mass can have sizable
couplings with the $K^*$ channel~\cite{CR98b}.
In addition, there are some preliminary experimental data from the CLAS
Collaboration at Jefferson Lab. on the reactions of $K^*$
photoproduction, i.e., $K^*\Sigma$ \cite{HH05b} and $K^*\Lambda$
\cite{GW06} production.
These experiments show that the total cross sections for $K^*$
photoproduction, though small, are not so much suppressed than those for
$K$ photoproduction,
and it leads to the conclusion that full coupled-channel analyses to
search for the resonances should include the
$K^*$ channel as well~\cite{GW06}.
Therefore, it is legitimate to study the production mechanisms of $K^*$
photoproduction.

At present, theoretical works to understand the $K^*$ photoproduction
reactions are very limited \cite{Thornber68}.
In Ref.~\cite{ZAB01}, Zhao et al. studied $K^* \Sigma$ photoproduction
from the proton targets using a quark model.
This model is based on the quark-meson couplings whose coupling constants
are assumed to be flavor-blind, which allows to use the values
determined by other reactions.
To implement the $t$-channel exchange contribution, the kaon exchange
was considered.
More accurate experimental data are needed to further test their model 
\cite{HH05b},
and the other channels for $K^*$ photoproduction like $K^*\Lambda$
were not considered.

In this paper, we investigate $K^*\Lambda$ photoproduction, $\gamma N
\to K^* \Lambda$.
The purpose of this work is to study the background production
mechanisms that include $t$-channel $K^*$, $K$, and $\kappa$ exchanges
as well as $s$-channel nucleon and $u$-channel hyperon
($\Lambda,\Sigma,\Sigma^*$) diagrams.
This can provide a platform for future investigation of nucleon resonances
that can also contribute to this reaction near the threshold.
Because of isospin, the $s$-channel $\Delta$ resonances are excluded, and
this reaction has an advantage in the study of nucleon resonances.
Our approach is based on the effective Lagrangians and similar to
the work of Ref.~\cite{OKL03d}.
By making use of the effective Lagrangians for $K^*$ meson interactions, we
evaluate the tree diagrams for $K^*$ photoproduction.
The coupling constants are constrained either by phenomenology or by 
quark model predictions when the experimental inputs are not available.
One advantage of $K^*$ photoproduction over $K$ photoproduction
is that it provides a chance to study the controversial scalar
$\kappa(700\mbox{---}900)$ meson \cite{PDG04} in the $t$-channel.
Such a contribution is prohibited in kaon photoproduction since
$\kappa \to K\gamma$ interaction is not allowed by angular momentum and
parity consideration.
We will see, however, that the $\kappa$ meson exchange is
suppressed in $K^*$ photoproduction and it would be hard to identify the
$\kappa$ meson contribution in this reaction at present.

Since both the $K^*$ and nucleon are isodoublets, we consider the following
two reactions,
\begin{equation}
\mbox{(I)}: \gamma p \to K^{*+} \Lambda, \qquad
\mbox{(II)}: \gamma n \to K^{*0} \Lambda.
\label{react}
\end{equation}
In the next Section, we develop our approach for $K^*$ photoproduction.
The effective Lagrangians and their coupling constants are discussed in
detail.
Our results for cross sections and some spin asymmetries are given in
Sec. III, and we make some comments on the Regge approach to this
reaction.
We summarize in Sec. IV.

\section{Model}

\begin{figure}
\centering
\epsfig{file=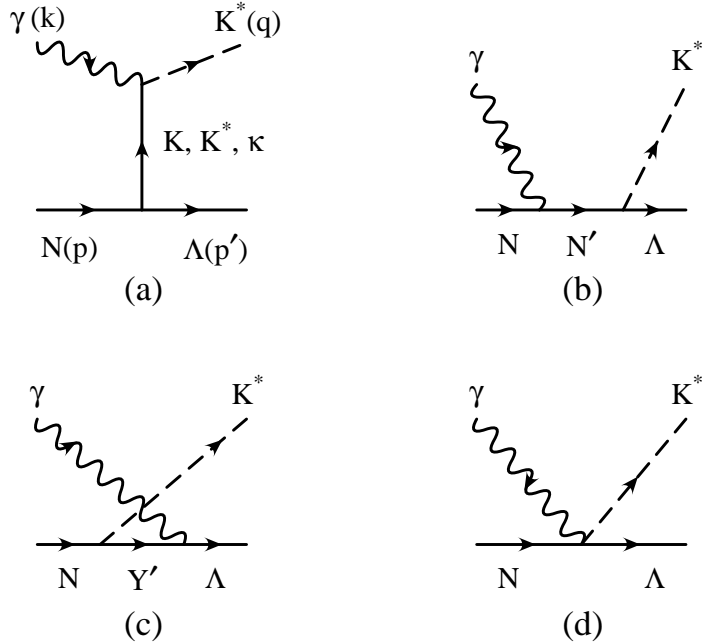, width=0.6\hsize}
\caption{Tree diagrams for $\gamma N \to K^* \Lambda$, which include (a)
$t$-channel exchanges, (b) intermediate nucleon, (c) intermediate
hyperon, and (d) contact diagrams.}
\label{fig:diag}
\end{figure}

The tree diagrams we are considering are shown in Fig.~\ref{fig:diag},
which also defines the momentum of each particle.
In this calculation, we work with a model which includes (i) $t$-channel
$K$, $K^*$, and $\kappa$ exchanges, (ii) $s$-channel nucleon,
and (iii) $u$-channel hyperon ($\Lambda,\Sigma,\Sigma^*$) terms.
The contact term for the charged $K^*$ photoproduction is included as well.
Because of isospin conservation, the $\Delta$ resonances cannot contribute
to this reaction.
The production amplitude can then be written as
\begin{equation}
\mathcal{M} = \varepsilon^*_\nu(K^*) \bar{u}_{\Lambda}(p') \mathcal{M}^{\mu\nu}
u_N(p) \varepsilon_\mu(\gamma),
\end{equation}
where $\varepsilon^\mu(K^*)$ and $\varepsilon^\mu(\gamma)$ are the
polarization vectors of $K^*$ vector meson and the photon, respectively.
The Dirac spinors of $\Lambda$ and the nucleon are denoted by
$u_{\Lambda}(p')$ and $u_N(p)$, respectively.
Below we calculate $\mathcal{M}^{\mu\nu}$ for each channel.

\subsection{$t$-channel $K^*$ and $K$ exchanges}

Because of charge, the $K^*$ exchange is present only for the charged $K^*$
photoproduction, $\gamma p \to K^{*+} \Lambda$.
The production amplitude is calculated from the following
effective Lagrangians,
\begin{eqnarray}
\mathcal{L}_{\gamma K^* K^*} &=& -ie A^\mu \left( K^{*-\nu}
K^{*+}_{\mu\nu} - K^{*-}_{\mu\nu} K^{*+\nu} \right),
\\
\mathcal{L}_{K^* N\Lambda} &=& - g_{K^* N\Lambda}^{} \overline{N} \left(
\gamma_\mu \Lambda
K^{*\mu} - \frac{\kappa_{K^* N\Lambda}^{}}{2M_N} \sigma_{\mu\nu} \Lambda
\partial^\nu K^{*\mu} \right) + \mbox{ H.c.},
\label{lag:KSNL}
\end{eqnarray}
where $A_\mu$ is the photon field, 
$K^{*\pm}_{\mu\nu} = \partial_\mu K^{*\pm}_\nu - \partial_\nu
K^{*\pm}_\mu$,
and the isodoublets are defined by
\begin{equation}
K^{*} = \left( \begin{array}{c} K^{*+} \\ K^{*0} 
\end{array} \right), \qquad
N = \left( \begin{array}{c} p \\ n \end{array} \right) .
\end{equation}
We use the following coupling constants determined by the Nijmegen potential
\cite{SR99-RSY99},
\begin{eqnarray}
&& g_{K^* N\Lambda} = -4.26, \qquad \kappa_{K^* N \Lambda} = 2.66
\qquad \mbox{(NSC97a)}, \nonumber \\
&& g_{K^* N\Lambda} = -6.11, \qquad \kappa_{K^* N \Lambda} = 2.43
\qquad \mbox{(NSC97f)}.
\end{eqnarray}

The production amplitude then reads
\begin{eqnarray}
\mathcal{M}_{K^*}^{\mu\nu} &=& \eta_{K^*}^{}
\frac{e}{(k-q)^2 - M_{K^*}^2}
\Gamma^{\mu\nu\alpha}_{K^*} (k,q) P_{\alpha\beta}(k-q)
\Gamma^\beta_{K^*N\Lambda}(q-k),
\end{eqnarray}
where $\eta_{K^*}^{} = 1$ and $0$ for the reaction (I) and (II) of
Eq.~(\ref{react}), respectively, and
\begin{eqnarray}
\Gamma^{\mu\nu\alpha}_{K^*} (k,q) &=& 2 q^\mu g^{\nu\alpha} - q^\alpha
g^{\mu\nu} + k^\nu g^{\mu\alpha}, 
\nonumber \\
P_{\alpha\beta}(k-q) &=& g_{\alpha\beta} - \frac{(k-q)_\alpha
(k-q)_\beta}{M_{K^*}^2},
\nonumber \\
\Gamma^\mu_{K^*N\Lambda}(q-k) 
&=& g_{K^* N\Lambda}^{} \left[ \gamma^\mu - \frac{i
\kappa_{K^*N\Lambda}^{}}{2M_N} \sigma^{\mu\nu} (q-k)_\nu \right].
\end{eqnarray}
The decay width of $K^*$, $\Gamma_{K^*} = 50.8$ MeV, is included by
replacing $M_{K^*}$ in the propagator by $M_{K^*} - i \Gamma_{K^*}/2$.

On the other hand, the $t$-channel kaon exchange is allowed for both
reactions.
In this case, we have
\begin{eqnarray}
\mathcal{L}_{\gamma K K^*} &=& g_{\gamma KK^*}^0
\varepsilon^{\mu\nu\alpha\beta} \partial_\mu A_\nu \left( \partial_\alpha
K_\beta^{*0} \bar{K}^0 + \partial_\alpha \bar{K}_\beta^{*0}
K^0 \right)
\nonumber \\ && \mbox{}
+ g_{\gamma KK^*}^c
\varepsilon^{\mu\nu\alpha\beta} \partial_\mu A_\nu \left( \partial_\alpha
K_\beta^{*-} K^+ + \partial_\alpha K_\beta^{*+} K^- \right),
\nonumber \\
\mathcal{L}_{KN\Lambda} &=& -i g_{KN\Lambda}^{} \overline{N} \gamma_5
\Lambda K + \mbox{ H.c.},
\label{lag2}
\end{eqnarray}
where $K$ is the kaon iso-doublet, $K^T = (K^+, K^0)$.
The coupling constants $g_{\gamma KK^*}^{}$ can be calculated from the
experimental data for $\Gamma(K^* \to K\gamma)$, which gives
\begin{equation}
g_{\gamma KK^*}^0 = -0.388 \mbox{ GeV}^{-1}, \qquad
g_{\gamma KK^*}^c = 0.254 \mbox{ GeV}^{-1},
\end{equation}
where the phases of the couplings are fixed from the quark model.

The coupling constant $g_{KN\Lambda}^{}$ is obtained by using SU(3)
flavor symmetry relation, which gives
\begin{equation}
g_{KN\Lambda}^{} = - \frac{1}{\sqrt3} (1+2f) g_{\pi NN} = -13.24,
\end{equation}
with $f = 0.365$ and $g_{\pi NN}^2/4\pi = 14.0$.
In this work we employ the pseudoscalar coupling for this interaction.
However, since the nucleon and $\Lambda$ are on their mass-shell, it is
equivalent to the pseudovector coupling.
The production amplitude for the $K$ exchange becomes 
\begin{equation}
\mathcal{M}^{\mu\nu}_{K} = 
\frac{i g_{\gamma KK^*}^{} g_{KN\Lambda}^{}}{(k-q)^2 -
M_K^2} \varepsilon^{\mu\nu\alpha\beta} k_\alpha q_\beta \gamma_5,
\end{equation}
where $g_{\gamma KK^*}^{} = g_{\gamma KK^*}^{c}$ for the reaction (I)
and $g_{\gamma KK^*}^{0}$ for the reaction (II).

\subsection{$t$-channel $\kappa$ exchange}

The scalar $\kappa$ meson cannot couple to $K\gamma$ because of angular
momentum and parity consideration and, as a result, the $\kappa$ meson
exchange is not present in kaon photoproduction.
However, $\gamma K^* \kappa$ coupling is allowed and this provides us
with a chance to study the controversial $\kappa(700\mbox{---}900)$ meson
\cite{Ishida04-Bugg04d} in $K^*$ photoproduction.

The effective Lagrangians for the scalar (and iso-doublet) $\kappa$ meson
interactions are given by
\begin{eqnarray}
\mathcal{L}_{\gamma K^*\kappa} &=& g_{\gamma K^*\kappa}^{}\,
A^{\mu\nu}\, \overline{\kappa} \, K^*_{\mu\nu} + \mbox{ H.c.},
\nonumber \\
\mathcal{L}_{\kappa N\Lambda} &=&
- g_{\kappa N \Lambda}^{}\,  \overline{N}\,  \kappa\,  \Lambda
+ \mbox{ H.c},
\end{eqnarray}
where $A_{\mu\nu} = \partial_\mu A_\nu - \partial_\nu A_\mu$ and
\begin{equation}
\kappa = \left( \begin{array}{c} \kappa^{+} \\ \kappa^{0} \end{array} \right),
\qquad
\overline{\kappa} = \left( \kappa^{-} ,  \bar{\kappa}^{0}  \right).
\end{equation}

The coupling constants are determined as follows.
For $g_{\gamma K^*\kappa}^{}$, we rely on the vector-meson dominance
model in the SU(3) limit~\cite{BHS02}. 
Here we briefly explain this model referring the details to
Ref.~\cite{BHS02}.
The basic idea of this model is to start with the most general
Lagrangian for the $SVV$ interaction, where $S$ stands for scalar meson
nonet and $V$ for vector meson nonet.
Then the $\bar{q}q$ or $\bar{q}^2q^2$ nature of scalar mesons is revealed
through the mixing angle between the scalar meson octet and scalar meson
singlet.
If the $\bar{q}q$ structure dominates the scalar meson wavefunction, then
one would expect the mixing angle $\theta_S \simeq -20^\circ$, while the
dominance of the tetraquark nature leads to
$\theta_S \simeq -90^\circ$ \cite{BHS02}.
The general form for the $SVV$ interaction can then be written as \cite{BHS02}
\begin{eqnarray}
\mathcal{L}_{SVV} &=& \beta_A^{} \epsilon_{abc} \epsilon^{a'b'c'}
[V_{\mu\nu}]^a_{a'} [V_{\mu\nu}]^b_{b'} S^c_{c'}
+ \beta_B^{} \mbox{Tr}\,(S) \mbox{Tr}\,(V^{\mu\nu} V_{\mu\nu})
\nonumber \\ && \mbox{}
+ \beta_C^{} \mbox{Tr}\,(SV^{\mu\nu}) \mbox{Tr}\,(V_{\mu\nu})
+ \beta_D^{} \mbox{Tr}\,(S) \mbox{Tr}\,(V^{\mu\nu})
\mbox{Tr}\,(V_{\mu\nu}).
\end{eqnarray}
Using the vector meson dominance hypothesis in the SU(3) limit,
the $SV\gamma$ couplings of our concern can be expressed in terms of the
above couplings $\beta_i$ and the mixing angle, and we have
\begin{eqnarray}
g_{\gamma K^*\kappa}^{c} = \frac{e}{g_\rho} \frac23 \beta_A, 
\qquad
g_{\gamma K^*\kappa}^{0} = -\frac{e}{g_\rho} \frac43 \beta_A, 
\end{eqnarray}
where $g_\rho = 4.04$ is the universal $\rho$ meson coupling,
and $g_{\gamma K^*\kappa}^{c} = g_{\gamma  K^{*-}\kappa^+}^{} =
g_{\gamma K^{*+}\kappa^- }^{}$ and $g_{\gamma K^*\kappa}^{0} =
g_{\gamma \bar{K}^{*0}\kappa^0}^{} = g_{\gamma K^{*0}\bar\kappa^0}^{}$.
Since $\beta_A$ is independent of the mixing angle $\theta_S$ \cite{BHS02},
this shows that the coupling constants $g_{\gamma K^*\kappa}^{}$ also do
not depend on the mixing angle and, therefore, they are
blind to whether the $\bar{q}q$ or $\bar{q}^2q^2$ nature dominates the
scalar meson structure in the SU(3) limit. 
Note also that the ratio of the couplings $g_{\gamma K^*\kappa}^{0}/g_{\gamma
K^*\kappa}^{c}$ is $-2$ in this limit just as in the case of $g_{\gamma
K K^*}^{0}/g_{\gamma KK^*}^c$, which is close to $-1.53$ in nature but
takes $-2$ in the SU(3) limit.
The coupling constant $\beta_A^{}$ can be estimated from the observed value
of $\Gamma(a_0 \to \gamma\gamma)$, which leads to $\beta_A^{} = 0.72$
GeV$^{-1}$ \cite{BHS02}.
Here we use $M(\kappa) = 900$ MeV and $\Gamma(\kappa) = 550$ MeV
following Ref.~\cite{BHS02}.%
\footnote{We note, however, that a recent analysis gives the pole
position of the $\kappa$ at $M = (750 \stackrel{+30}{\mbox{\scriptsize$-55$}})
- i (342 \pm 60)$ MeV \cite{Bugg06}.}

For the couplings of scalar mesons with octet baryons, we again use the
values of the Nijmegen potential \cite{SR99-RSY99}, which gives
\begin{eqnarray}
&& g_{\kappa N\Lambda}^{} \sim -8.3  \qquad \mbox{(NSC97a)},
\nonumber \\
&& g_{\kappa N\Lambda}^{} \sim -10.0  \qquad \mbox{(NSC97f)}.
\end{eqnarray}
However, it should be mentioned that the above values are obtained with
$M(\kappa) = 880$ MeV.
Also in Ref.~\cite{SR99-RSY99}, it was stressed that the structure of the
scalar mesons is crucial for the central $YN$ potential and the above values
are obtained with assuming that the scalar mesons are close to $\bar{q}q$
state.
With this caveat in mind,  we use the above values just as a guide for the
couplings involving the $\kappa$ meson.
Our numerical results show that the $\kappa$ meson exchange is
suppressed and the uncertainties of $\kappa$ meson coupling constants
are not crucial in $K^*$ photoproduction.
Collecting the $\kappa$ meson coupling constants, we have
\begin{eqnarray}
|g_{\gamma K^*\kappa}^{c} g_{\kappa N \Lambda}| &=& (1.0 \sim 1.2)\, e \mbox{
GeV}^{-1},
\nonumber \\
g_{\gamma K^*\kappa}^{0} g_{\kappa N \Lambda} &=&  -2
g_{\gamma K^*\kappa}^{c} g_{\kappa N \Lambda}.
\label{coup:kappa}
\end{eqnarray}
In fact, the phase of $g_{\gamma K^*\kappa}^{c} g_{\kappa N \Lambda}$
cannot be fixed at this stage.
However, since the $\kappa$ exchange contribution is small, the phase of
the above coupling constants is hard to be distinguished in $K^*$
photoproduction.
The production amplitude reads
\begin{equation}
\mathcal{M}^{\mu\nu} = - \frac{2 g_{\gamma K^*\kappa}^{} g_{\kappa N
Y}^{}}{(k-q)^2 - (M_\kappa-i\Gamma_\kappa/2)^2}
(k \cdot q g^{\mu\nu} - k^\nu q^\mu ),
\end{equation}
where $g_{\gamma K^*\kappa}^{} = g_{\gamma K^*\kappa}^{c}$ for the
reaction (I) and $g_{\gamma K^*\kappa}^{0}$ for the reaction (II).

\subsection{$s$-channel diagrams}

The $s$-channel diagrams shown in Fig.~\ref{fig:diag}(b)
can contain the intermediate nucleon  as well as nucleon resonances.
The purpose of this work is to investigate the main production
mechanisms which should be well understood before studying the nucleon
resonances.
In this work, therefore, we consider only the intermediate nucleon state 
postponing the inclusion of nucleon resonances to a future study 
as it requires more information or assumptions.
The consequences of limiting the intermediate state to the nucleon will be
discussed later.

The amplitude of the $s$-channel nucleon term can be calculated from
$\mathcal{L}_{K^*N\Lambda}$ of Eq.~(\ref{lag:KSNL}) and
\begin{equation}
\mathcal{L}_{\gamma NN} = -e \overline{N} \left[ \gamma_\mu A^\mu
\frac{1+\tau_3}{2} - \frac{1}{2M_N} ( \kappa_s^N + \kappa_v^N \tau_3)
\sigma_{\mu\nu} \partial^\nu A^\mu \right] N,
\end{equation}
where the isoscalar and isovector anomalous magnetic moments of the
nucleon are $\kappa_s^N = -0.06$ and $\kappa_v^N = 1.85$.
Then the production amplitude is obtained as
\begin{equation}
\mathcal{M}^{\mu\nu}_{N} =
\frac{e}{(k+p)^2 - M_N^2} \Gamma^\nu_{K^*N\Lambda}(q) (k\!\!\!/ + p\!\!\!/ +
M_N) \Gamma^\mu_{\gamma N}(k),
\end{equation}
where
\begin{equation}
\Gamma^\mu_{\gamma N}(k) = \gamma^\mu Q_N  + \frac{i \kappa_N^{}}{2M_N}
\sigma^{\mu\nu} k_\nu,
\end{equation}
with ($Q_p = +1$, $\kappa_p = 1.79$) for the reaction (I)
and ($Q_n = 0$, $\kappa_n = -1.91$) for the reaction (II).

\subsection{$u$-channel diagrams}

For the $u$-channel diagrams of Fig.~\ref{fig:diag}(c), we consider
intermediate hyperons including $\Lambda(1116)$, $\Sigma(1193)$, and
$\Sigma^*(1385)$.
The diagrams with the intermediate octet hyperons can be calculated with
\begin{eqnarray}
\mathcal{L}_{\gamma\Lambda\Lambda} &=& \frac{e\kappa_\Lambda^{}}{2M_N}
\overline{\Lambda} \sigma_{\mu\nu} \partial^\nu A^\mu \Lambda,
\\
\mathcal{L}_{\gamma\Sigma\Lambda} &=& \frac{e\mu_{\Sigma\Lambda}^{}}{2M_N}
\overline{\Sigma}^0 \sigma_{\mu\nu} \partial^\nu A^\mu \Lambda
+ \mbox{
H.c.},
\end{eqnarray}
where $\kappa_\Lambda^{} = -0.61$ and $\mu_{\Sigma\Lambda}^{} = 1.62 \pm 0.08$.
This leads to
\begin{eqnarray}
\mathcal{M}^{\mu\nu}_\Lambda &=& \eta_\Lambda^{}
\frac{e}{(p-q)^2 - M_\Lambda^2}
\Gamma^\mu_{\gamma\Lambda}(k) (p\!\!\!/ - q\!\!\!/ + M_\Lambda)
\Gamma^\nu_{K^*N\Lambda}(q),
\nonumber \\
\mathcal{M}^{\mu\nu}_\Sigma &=& \eta_\Sigma^{}
\frac{e}{(p-q)^2 - M_\Sigma^2}
\Gamma^\mu_{\Sigma\Lambda}(k) (p\!\!\!/ - q\!\!\!/ + M_\Sigma)
\Gamma^\nu_{K^*N\Sigma}(q),
\end{eqnarray}
where
\begin{eqnarray}
\Gamma^\mu_{\gamma\Lambda} (k) &=& \frac{i \kappa_\Lambda^{}}{2M_N}
\sigma^{\mu\nu} k_\nu,
\nonumber \\
\Gamma^\mu_{\Sigma\Lambda} (k) &=& \frac{i \mu_{\Sigma\Lambda}^{}}{2M_N}
\sigma^{\mu\nu} k_\nu,
\end{eqnarray}
with $\eta_\Lambda^{} = 1$ for the reactions (I) and (II),
and $\eta_\Sigma^{} = 1$, $-1$ for the reaction (I) and (II),
respectively, which comes from the isospin factors.
The vertex function $\Gamma^\nu_{K^*N\Lambda}(q)$ was given before and
$\Gamma^\nu_{K^*N\Sigma}(q)$ has the same structure but with \cite{SR99-RSY99}
\begin{eqnarray}
&& g_{K^* N\Sigma}^{} = -2.46, \qquad \kappa_{K^* N \Sigma}^{} = -0.47
\qquad \mbox{(NSC97a)}, \nonumber \\
&& g_{K^* N\Sigma}^{} = -3.52, \qquad \kappa_{K^* N \Sigma}^{} = -1.14
\qquad \mbox{(NSC97f)}.
\end{eqnarray}

In order to compute the contribution from the intermediate
$\Sigma^*(1385)$, we need to know the interactions
$\mathcal{L}_{K^* N \Sigma^*}$ and $\mathcal{L}_{\gamma \Lambda\Sigma^*}$.
The general form for $\mathcal{L}_{K^* N \Sigma^*}$ is written as
\begin{eqnarray}
\mathcal{L}_{K^*N\Sigma^*} &=&
-i \frac{f^{(1)}_{K^*N\Sigma^*}}{M_{K^*}} \overline{K^*}_{\mu\nu}
\overline{\bm{\Sigma}^*}^{\mu} \cdot \bm{\tau} \gamma^\nu \gamma_5 N
- \frac{f^{(2)}_{K^*N\Sigma^*}}{M_{K^*}} \overline{K^*}_{\mu\nu}
\overline{\bm{\Sigma}^*}^{\mu} \cdot \bm{\tau} \gamma_5 \partial^\nu N
\nonumber \\ && \mbox{}
+ \frac{f^{(3)}_{K^*N\Sigma^*}}{M_{K^*}} \partial^\nu
\overline{K^*}_{\mu\nu} \overline{\bm{\Sigma}^*}^{\mu} \cdot \bm{\tau}
\gamma_5 N + \mbox{ H.c.},
\label{K*NS*}
\end{eqnarray}
which follows from the fact that this is an interaction of
$J^P = \frac32^+ \to \frac12^+ + 1^-$.
Thus we have, in general, three independent couplings.
However their values are poorly known and we use the SU(3)
symmetry relations to estimate the couplings.
(See, e.g., Ref.~\cite{OK04}.)
By making use of the quark model prediction and
SU(3) flavor symmetry we obtain
\begin{equation}
f^{(1)}_{K^*N\Sigma^*} = -\frac{1}{\sqrt6} \frac{M_{K^*}}{M_\rho}
f^{(1)}_{\rho N\Delta} = -2.6,
\end{equation}
with $f^{(1)}_{\rho N\Delta} = 5.5$ \cite{KA04,ONL04}.
The other couplings are unknown and we do not consider the terms
containing $f^{(2)}_{K^*N\Sigma^*}$ and $f^{(3)}_{K^*N\Sigma^*}$ \cite{ONL04}.

The Lagrangian for $\gamma \Lambda \Sigma^*$ interaction has the same
structure as $\mathcal{L}_{K^*N\Sigma^*}$ of Eq.~(\ref{K*NS*}).
Since the photon is massless, the number of independent couplings is
reduced to 2 and the interaction can be written as
\begin{eqnarray}
\mathcal{L}_{\gamma \Lambda \Sigma^*} = \frac{ieg_1}{2M_N}
\overline{\Sigma^*}_\mu
\gamma_\nu \gamma_5 \Lambda F^{\mu\nu}
+ \frac{eg_2}{4M_N^2} \overline{\Sigma^*}_\mu \gamma_5
\partial_\nu \Lambda F^{\mu\nu} + \mbox{ H.c.},
\label{eq:tran1}
\end{eqnarray}
which leads to the decay width as
\begin{eqnarray}
\Gamma(\Sigma^* \to \Lambda\gamma) &=&
\frac{p_\gamma^3}{48\pi M_{\Sigma^*}^2} \left(
\frac{e}{2M_N} \right)^2 \Biggl\{ \left[ g_{1} (3 M_{\Sigma^*} +
M_\Lambda) - g_{2}^{} \frac{M_{\Sigma^*}}{2M_N} (M_{\Sigma^*} -
M_\Lambda) \right]^2
\nonumber \\ && \mbox{} \qquad\qquad\qquad
+ 3 \left[ g_{1}^{} - g_{2}^{} \frac{M_{\Sigma^*}}{2M_N}
\right]^2 (M_{\Sigma^*} - M_\Lambda )^2 \Biggr\},
\end{eqnarray}
and the $E2/M1$ ratio as \cite{NBL90}
\begin{equation}
R_{EM} = E2/M1 = -\frac{M_{\Sigma^*} - M_\Lambda}{2M_N} \frac{g_1 - g_2
M_{\Sigma^*}/(2M_N)}
{g_1(3M_{\Sigma^*} + M_\Lambda)/(2M_N) - g_2 M_{\Sigma^*}
(M_{\Sigma^*}-M_\Lambda)/(2M_N)^2}.
\end{equation}

The recent CLAS experiment puts a constraint on the radiative decay width
of $\Gamma(\Sigma^*\to\Lambda\gamma)$ as \cite{CLAS05a}
\begin{equation}
\Gamma(\Sigma^*\to\Lambda\gamma) = 479 \pm 120
\stackrel{+81}{\mbox{\scriptsize $-100$}}
\mbox{ keV}.
\end{equation}
Together with the chiral quark model prediction on the $E2/M1$ ratio
for this radiative decay, $R_{EM} = -2.0$ \% \cite{WBF98}, we obtain 
\begin{equation}
g_1 = 3.78, \qquad g_2 = 3.18.
\end{equation}
The production amplitude reads
\begin{equation}
\mathcal{M}^{\mu\nu}_{\Sigma^*} = \eta_{\Sigma^*}^{}
\frac{e}{(p-q)^2-M_{\Sigma^*}^2} \Gamma^{\mu\beta}_{\Sigma^* \Lambda} (k,p')
\Delta_{\beta\alpha}(\Sigma^*, p-q) \Gamma^{\nu\alpha}_{K^* N
\Sigma^*}(q),
\end{equation}
where $\eta_{\Sigma^*}^{} = 1$ for the reaction (I),
$\eta_{\Sigma^*}^{} = -1$ for the reaction (II), and
\begin{eqnarray}
\Gamma^{\nu\alpha}_{K^* N\Sigma^*}(q) &=&
\frac{f^{(1)}_{K^*N\Sigma^*}}{M_{K^*}} \gamma_\delta \gamma_5 (q^\alpha
g^{\nu\delta} - q^\delta g^{\nu\alpha}),
\nonumber \\
\Gamma^{\mu\beta}_{\Sigma^* \Lambda} (k,p') &=& \left\{ \frac{g_1}{2M_N}
\gamma_\nu \gamma_5 + \frac{g_2}{4M_N^2} p_\nu' \gamma_5 \right\} \left(
k^\beta g^{\mu\nu} - k^\nu g^{\mu\beta} \right).
\end{eqnarray}
The spin-3/2 Rarita-Schwinger propagator for the
resonance $R$ with momentum $p$ contains
\begin{equation}
\Delta_{\mu\nu}(R,p) = (p\!\!\!/ + M_R)
\left(-g_{\mu\nu} + \frac13 \gamma_\mu \gamma_\nu +
\frac{1}{3M_R} (\gamma_\mu p_\nu - \gamma_\nu p_\mu) + \frac{2}{3M_R^2} p_\mu
p_\nu \right) .
\end{equation}
The decay width is incorporated by replacing
$M_R \rightarrow M_R - i \Gamma_R/2$ in the propagator.
We use $M_{\Sigma^*} = 1385$ MeV and $\Gamma_{\Sigma^*}
= 37$ MeV.

\subsection{Contact diagram}

Since the $K^* N \Lambda$ interaction contains a derivative coupling,
there exists a contact diagram for the charged $K^*$ vector meson 
photoproduction.
Inclusion of this diagram is essential to satisfy the gauge-invariance
condition.
By minimal substitution in the Lagrangian (\ref{lag:KSNL}), we have
\begin{eqnarray}
\mathcal{L}_{\gamma K^*N\Lambda} &=& -i
\frac{eg_{K^*N\Lambda}^{}\kappa_{K^*N\Lambda}^{}}{2M_N}
\overline{\Lambda} \sigma^{\mu\nu} A_\nu K^{*-}_\mu p + \mbox{ H.c.},
\end{eqnarray}
which gives the contact diagram of Fig.~\ref{fig:diag}(d).
The corresponding amplitude is given by  
\begin{equation}
\mathcal{M}_C^{\mu\nu} = - 
\frac{ieg_{K^*N\Lambda}^{}\kappa_{K^*N\Lambda}^{}}{2M_N} \sigma^{\mu\nu}.
\end{equation}

\subsection{Form factors}

The form factors are included to dress the vertices of the diagrams.
For the form factors of $t$-channel exchanges, $F_{K^*}$, $F_K$, and
$F_\kappa$, we use the form of
\begin{equation}
F_M(p^2) = \frac{\Lambda^2 - M^2_{\rm ex}}{\Lambda^2 - p^2},
\label{FF:mono}
\end{equation}
where $M^2_{\rm ex}$ and $p^2$ are the mass and momentum squared of the
exchanged particle.
The form factor is multiplied to each vertex, and each diagram contains two
powers of the form factor.

The $s$- and $u$-channel diagrams have the form factor, $F_N$,
$F_\Lambda$, $F_\Sigma$, and $F_{\Sigma^*}$, in the form of \cite{PJ91}
\begin{equation}
F_B(p^2) = \left( \frac{n\Lambda^4}{n\Lambda^4 + (p^2-M_{\rm ex}^2)^2}
\right)^n,
\label{ff:PJ}
\end{equation}
which becomes the Gaussian form as $n \to \infty$.
We take $n=1$ but the results with $n \to \infty$ will also be
discussed.

It is well-known that introducing the form factors that depend on the
momentum and mass of the exchanged particle violates the charge 
conservation condition, $k_\mu
\mathcal{M}^{\mu\nu} = 0$ unless the production amplitude is transverse
by itself.
For example, in the reaction of $\gamma p \to K^{*+} \Lambda$, 
the $t$-channel $K^*$ exchange, $s$-channel nucleon term, and the
contact term separately violate the charge-conservation condition
but their sum does not.
Having different form factor at each channel clearly makes the sum
violate the charge conservation.
Various methods to restore the charge-conservation condition have been
developed \cite{Ohta89,Habe97,HBMF98a,DW01a}.
In this work, following Ref.~\cite{DW01a},
we take the common form factor for the $t$-channel $K^*$ exchange, $s$-channel
nucleon term, and the contact term as
\begin{equation}
F = 1 - (1 - F_{K^*})(1-F_N).
\label{ff}
\end{equation}

In the case of $\gamma n \to K^{*0} \Lambda$, each production amplitude
is transverse.
Thus the charge-conservation condition is satisfied even with the form
factors and no prescription like Eq.~(\ref{ff}) is necessary.

\section{Results}

\begin{figure}
\centering
\epsfig{file=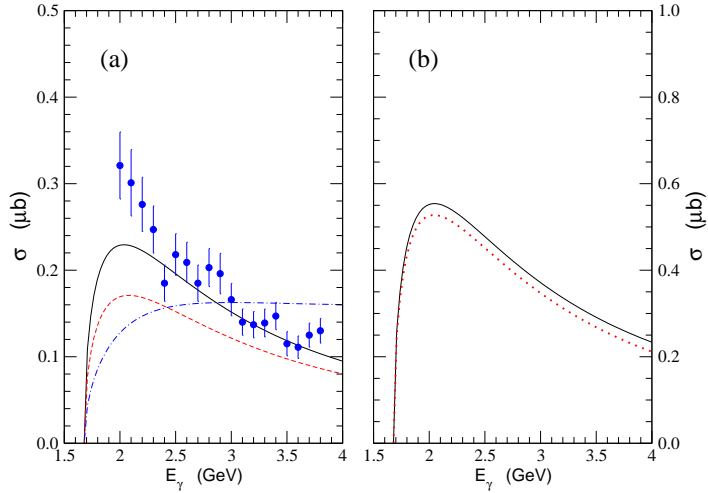, width=0.5\hsize, angle=-90}
\caption{(color online).
Total cross section (a) for $\gamma p \to K^{*+} \Lambda$
and (b) for $\gamma n \to K^{*0} \Lambda$. In (a), the solid line is obtained
with $\Lambda_{K^*} = 0.9$ GeV, while the dashed and dot-dashed lines 
are obtained with $\Lambda_{K^*} = 1.0$ and $1.1$ GeV, respectively.
In (b), the solid line is the full calculation and the dotted line 
is obtained from the $t$-channel $K$ exchange alone.
The experimental data are from Ref.~\cite{GW06}.}
\label{fig:total}
\end{figure}

Before we present our numerical results, the cutoff parameters should be
fixed.
We use the total cross section for $\gamma p \to K^{*+} \Lambda$
reported in Ref. \cite{GW06} to constrain the cutoff parameters of the
form factors.
The observed total cross section data show that the cross section has
the maximum near the threshold and then decreases as the energy increases.
This behaviour is observed in the model of spin-0 meson
exchanges, while the spin-1 meson exchange makes the total cross section
increase with the energy since the total cross section in the
$t$-channel exchange model scales as $\sigma \sim s^{J-1}$, where $J$ is
the spin of the exchanged particle in $t$-channel.
In our case, the charged $K^*$ production contains the $K^*$ vector
meson exchange and, as a result, it gives an increasing total cross
section with the energy.
This is shown by the dot-dashed line of Fig.~\ref{fig:total}(a), which is
obtained with the cutoff $\Lambda_{K^*} = 1.1$ GeV.
(See below for the other cutoff parameters.)
However, this is not consistent with the experimental observation
reported by Ref. \cite{GW06}, which means that the $K^*$ exchange
contribution should be suppressed.
In fact, the contribution from the higher-spin meson exchanges can be
modified by reggeizing the production amplitude. (See Sec.~III.)
In this exploratory work, however, in order to avoid additional
complexity, we simply suppressed the $K^*$ exchange by employing a soft
form factor with $\Lambda_{K^*} = 0.9$ GeV.

Shown in Fig.~\ref{fig:total} are the total cross sections for $\gamma p
\to K^{*+} \Lambda$ (left panel) and $\gamma n \to K^{*0} \Lambda$
(right panel).
These results are obtained with the (NSC97a) values for the $K^*$ and
$\kappa$ couplings.
The solid lines are obtained with $\Lambda_{K^*} = 0.9$ GeV and $\Lambda_K
= \Lambda_\kappa = 1.1$ GeV, while the $s$- and $u$-channel form factors
have $\Lambda = 0.9$ GeV \cite{ONL04}.
This gives a good fit to the measured total cross sections except the
near-threshold region.
In the left panel we also give the results obtained with $\Lambda_{K^*} = 1.0$
GeV (dotted line) and with $\Lambda_{K^*} = 1.1$ GeV (dot-dashed line)
while keeping the other cutoff parameters.
The decrease of the total cross section by changing $\Lambda_{K^*} =
0.9$ GeV to $\Lambda_{K^*} = 1.0$ GeV shows the destructive interference
between the $K$ exchange and $K^*$ exchange.
With $\Lambda_{K^*} = 1.1$ GeV, the vector meson exchange starts to
dominate and the total cross section shows the behavior expected from the
vector meson exchange model.
With $\Lambda_{K^*} = 0.9$ GeV, the $K^*$ vector meson exchange is suppressed
and, in fact, the kaon exchange dominates the reaction.

The $K$ meson exchange dominance can be also seen in the neutral $K^*$
production shown in Fig.~\ref{fig:total}(b).
The vector meson exchange does not contribute to this reaction 
and Fig.~\ref{fig:total}(b) shows the behavior expected from the 
pseudoscalar meson exchange model.
One can clearly see from the dotted line that the cross section
is almost dominated by the $K$ exchange. 
Figure~\ref{fig:total} also shows that the cross section for the neutral
$K^*$ photoproduction is larger than that for the charged $K^*$
photoproduction.
This can be understood by the dominance of $K$ exchange and the ratio of
$|g_{\gamma KK^*}^0/g_{\gamma KK^*}^c| \simeq 1.53$.

A close inspection of our results for total cross sections with the data of
Ref.~\cite{GW06} shows that our model can describe the charged $K^*$
meson production process at large energies, $E_\gamma > 2.3$ GeV.
But there is discrepancy between the two at lower energies.
This may be ascribed to limiting the $s$- and $u$-channel diagrams to
the intermediate lowest octet and decuplet baryons.
We expect that the low energy behavior can be improved by including the
nucleon resonances lying near the $K^*\Lambda$ threshold.

\begin{figure}
\centering
\epsfig{file=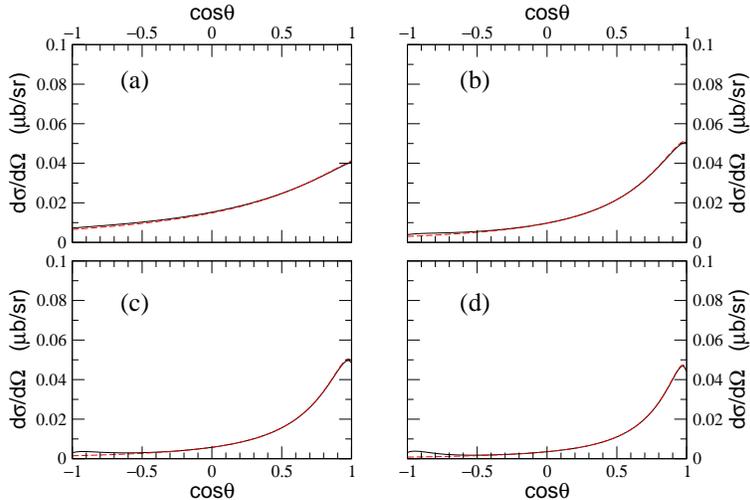, width=0.5\hsize, angle=-90}
\caption{(color online).
Differential cross sections for $\gamma p \to K^{*+} \Lambda$
at $E_\gamma = $ (a) 2.0 GeV, (b) 2.5 GeV, (c) 3.0 GeV, and (d) 3.5 GeV.
The solid lines are the full calculation and dashed lines are for the
$K$ meson exchange alone.}
\label{fig:dif-p}
\end{figure}

\begin{figure}
\centering
\epsfig{file=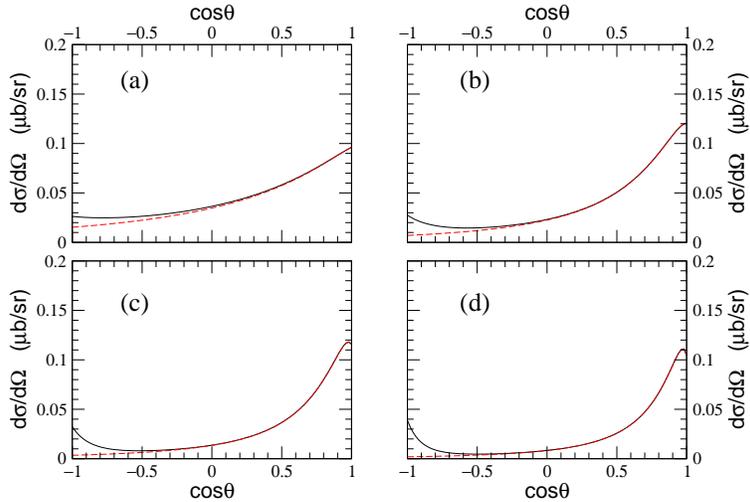, width=0.5\hsize, angle=-90}
\caption{(color online).
Differential cross sections for $\gamma n \to K^{*0} \Lambda$
at $E_\gamma = $ 2.0 GeV, (b) 2.5 GeV, (c) 3.0 GeV, and (d) 3.5 GeV.
The solid lines are the full calculation and dashed lines are for the
$K$ meson exchange alone.}
\label{fig:dif-n}
\end{figure}

The differential cross sections for the charged and neutral $K^*$ 
photoproduction are given in Figs.~\ref{fig:dif-p} and
\ref{fig:dif-n}, respectively, at four photon energies, $E_\gamma
= 2.0$, $2.5$, $3.0$, and $3.5$ GeV.
They are given as functions of the scattering angle $\theta$, which is
defined as the angle between the photon beam and the outgoing $K^*$
vector meson in the center-of-mass frame.
In both cases, we have a forward peak as a result of 
the $K$-exchange dominance.
The effects of the other production amplitudes can be barely seen 
only at large scattering angle region.
The contribution coming from the scalar $\kappa$ meson exchange is
suppressed in the considered energy region.
We have varied the $\kappa$ meson couplings including the phase around
the values of Eq.~(\ref{coup:kappa}) with the form factor
(\ref{FF:mono}) in the production amplitude, but the changes are not
crucial.

We have also employed the Gaussian form factor by taking the limit of $n \to
\infty$ in Eq.~(\ref{ff:PJ}) and found 
no significant difference in the differential cross sections. 
\footnote{Note, in some other reactions, 
the difference between $n=1$ and $n \to \infty$ becomes quite
noticeable, e. g., in Ref.~\cite{OSL01-OSLW02}.}
This is because of the $K$ meson exchange dominance.
The only difference could be seen in the backward scattering region,
$\cos\theta \le -0.5$, because the form factor with $n \to \infty$
suppresses the differential cross section at large scattering angles
more than that with $n=1$.
However, it is hard to distinguish them.

\begin{figure}
\centering
\epsfig{file=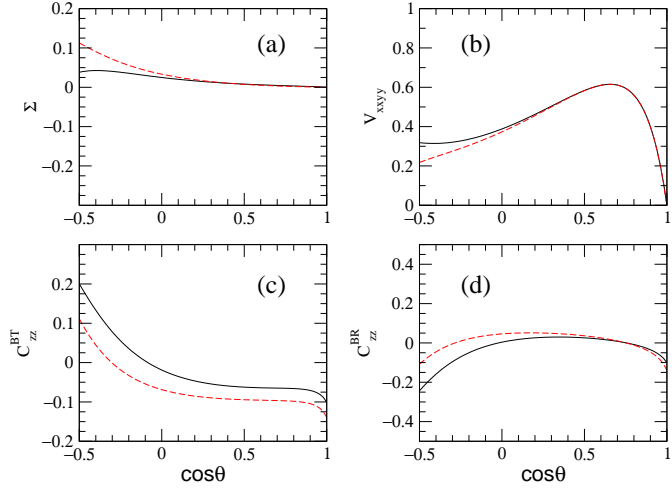, width=0.5\hsize, angle=-90}
\caption{(color online).
Spin asymmetries (a) $\Sigma$, (b) $V_{xxyy}$, (c) $C_{zz}^{BT}$, and
(d) $C_{zz}^{BR}$ for $\gamma p \to K^{*+}\Lambda$ (solid lines) and for
$\gamma n \to K^{*0} \Lambda$ (dashed lines) at $E_\gamma = 3.0$ GeV.}
\label{fig:pol}
\end{figure}

Next we consider the spin asymmetries.
Since contributions from the baryon resonances are expected to be seen
mostly in the backward scattering angles, $\cos\theta<0$,
here we focus on the spin asymmetries in the range $\cos\theta \ge -0.5$.
Because of the $K$-exchange dominance, the single asymmetries for photon,
nucleon, and $\Lambda$ are close to zero for forward scattering angles.
As an example, in Fig.~\ref{fig:pol}(a), the single photon asymmetry is
shown which is defined as
\begin{equation}
\Sigma = \frac{\sigma^\parallel - \sigma^\perp}{\sigma^\parallel +
 \sigma^\perp},
\end{equation}
where $\sigma^{\parallel}$ ($\sigma^\perp$) is the differential cross section
produced by a photon linearly polarized along the $\hat{\bf x}$ and
($\hat{\bf y}$) axis in the CM frame.
Also given in Fig.~\ref{fig:pol} are the tensor polarization asymmetry
$V_{xxyy}$ of the $K^*$ vector meson, beam-target double asymmetry
$C_{zz}^{\rm BT}$, and beam-recoil double asymmetry $C_{zz}^{\rm BR}$.
The solid lines in Fig.~\ref{fig:pol} are the results for $\gamma p \to
K^{*+} \Lambda$ and the dashed lines for $\gamma n \to K^{*0} \Lambda$. 
The definition for these asymmetries and the coordinate system can be
found, e.g., in Refs.~\cite{TOYM98}.

\begin{figure}
\centering
\epsfig{file=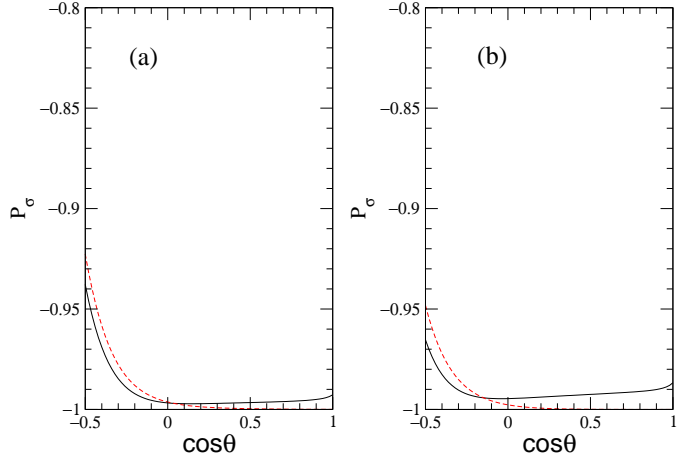, width=0.5\hsize, angle=-90}
\caption{(color online).
Parity asymmetry $P_\sigma$ (a) for $\gamma p \to K^{*+}\Lambda$ 
and (b) for $\gamma n \to K^{*0} \Lambda$ at $E_\gamma = 3.0$ GeV.
The solid lines are the full
calculation while the dashed lines are obtained without the $\kappa$
exchange.}
\label{fig:psigma}
\end{figure}

Another interesting feature of $K^*$ production reaction is that the scalar
$\kappa$ meson exchange is allowed.
In fact, since the $\kappa$ exchange is a natural-parity exchange,
while the $K$ meson exchange is an unnatural-parity exchange, the relative
strength of them can be estimated by measuring the parity asymmetry
defined as \cite{CSM68,SSW70}
\begin{equation}
P_\sigma \equiv \frac{\sigma^N - \sigma^U}{\sigma^N + \sigma^U}
= 2 \rho_{1-1}^1 - \rho^1_{00},
\end{equation}
where $\sigma^N$ and $\sigma^U$ are the contributions of
natural and unnatural parity exchanges to the cross section, and the
asymmetry $P_\sigma$ can be expressed in terms of the $K^*$ density matrix
elements.
{}From its definition, it can be easily found that $P_\sigma \to -1$ for
the $K$ exchange, while it becomes $+1$ for $K^*$ and $\kappa$ exchanges.
The reaction $\gamma n \to K^{*0} \Lambda$, where the $K^*$ exchange 
does not contribute, is a good place to estimate the relative strength
between the $K$ exchange and the $\kappa$ exchange.
In our model, however, the contribution from the $\kappa$ exchange is
suppressed and $P_\sigma$ is very close to $-1$ as can be seen from
Fig.~\ref{fig:psigma}. 
(The photon polarization asymmetry $\Sigma_V$ \cite{SSW70} also gives
the similar results.)

\begin{figure}
\centering
\epsfig{file=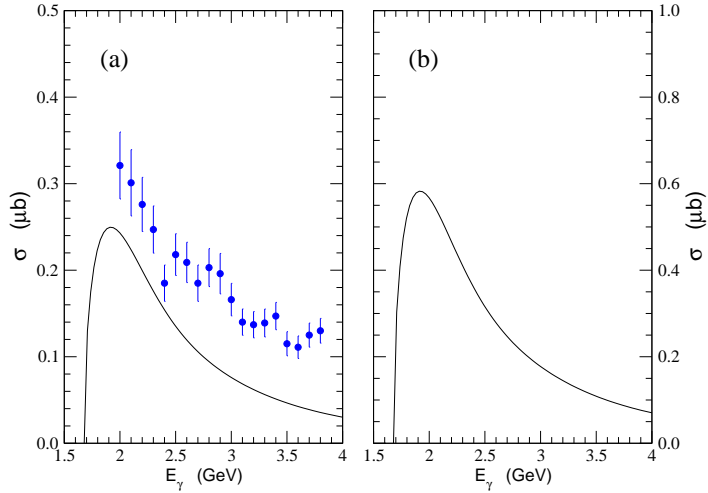, width=0.5\hsize, angle=-90}
\caption{(color online).
Total cross section (a) for $\gamma p \to K^{*+} \Lambda$
and (b) for $\gamma n \to K^{*0} \Lambda$ within $K$-trajectory exchange
model of Regge approach. 
The experimental data are from Ref.~\cite{GW06}.}
\label{fig:total-rg}
\end{figure}

Before closing, we briefly mention about the Regge approach to this
reaction.
In order to estimate the cross sections of $K^*\Lambda$ production in
Regge approach, we have employed the method of Ref. \cite{GLV97} to
reggeize the amplitudes, namely, the propagator is replaced by the Regge
propagator while keeping the coupling constants as before.
The results within the $K$-trajectory exchange process are presented in
Fig.~\ref{fig:total-rg}.
Although it underestimates the total cross sections for $K^{*+}\Lambda$
production, this shows that the energy dependence of the total cross
sections is similar to the case of the tree-level approximation that is
employed in this work.
We refrain from making the full calculations including the
$K^*$-trajectory exchange and other possible contributions, which is beyond
the scope of this work.
However, we expect that the discrepancies shown in
Fig.~\ref{fig:total-rg} could be compensated by the $K^*$-trajectory
exchange, which will dominate in the high energy region, as claimed by
Ref.~\cite{Thornber68}, because the $K^*$ trajectory lies above the $K$
trajectory: $\alpha_{K^*}^{}(t) \approx 0.25 + 0.83\, t$ and
$\alpha_K^{}(t) \approx -0.17 + 0.7\, t$ \cite{GLV97}.

\section{Summary}

We have investigated photoproduction mechanisms of $K^* \Lambda$ 
off the nucleon targets.
Our calculation includes the $t$-channel strange meson exchanges 
as well as the $s$- and $u$-channel intermediate nucleon and hyperon diagrams.
Baryon resonances, which are predicted to have sizable couplings
with $K^*\Lambda$~\cite{CR98b}, can also participate in the reaction.
Our calculation thus can provide background production mechanisms to
investigate such resonances in $K^* \Lambda$ photoproduction. 
We have found that the $K$-meson exchange dominates both the charged 
and neutral $K^*$ photoproduction, which leads to sharp peaks in the
differential cross sections at forward scattering angles.
Because of the $K$-exchange dominance, the total cross sections for
the neutral $K^*$ photoproduction is found to be larger than those for
the charged $K^*$ photoproduction.
Comparison with the experimental data of Ref.~\cite{GW06} shows that
the inclusion of baryon resonances can improve our model prediction at
lower energies close to the threshold.

As a test for our model, we have made several predictions
on the spin asymmetries which can be measured at current experimental
facilities.
One advantage of $K^*$ photoproductions over $K$ photoproductions is
that it allows the $\kappa$-meson exchange whose existence and
properties are still under debate.
However, within our model, we found that the contribution from the
$\kappa$ meson exchange is suppressed and can hardly be seen in the
reaction of $K^* \Lambda$ photoproduction.
Therefore, the parity asymmetry $P_\sigma$, which can distinguish the relative
strength between the $\kappa$ and $K$ exchanges especially in the neutral
$K^*$ production, is found to be $P_\sigma \simeq -1$
due to the $K$-meson exchange dominance. 
Measurement of those spin asymmetries would be helpful to test the
reaction mechanisms of $K^*$ photoproduction such as the
dominance of the $K$ meson exchange.

\acknowledgments

We are grateful to L. Guo, K. Hicks, T.-S.H. Lee, and K.~Nakayama for
fruitful discussions and encouragements.
We also thank L. Guo and D. P. Weygand for providing us with the preliminary
data for $K^* \Lambda$ photoproduction.
Y.O. was supported by Forschungszentrum-J{\"u}lich, contract
No. 41445282 (COSY-058).

\end{document}